\documentclass[11pt]{article}
\usepackage[T1]{fontenc}
\usepackage{lmodern}
\usepackage[a4paper,margin=1in]{geometry}
\usepackage{amsmath,amssymb,amsthm}
\usepackage{graphicx}
\usepackage[colorlinks=true,allcolors=blue]{hyperref}
\usepackage{natbib}
\usepackage{xspace}
\usepackage{ulem}
\usepackage{xcolor}
\usepackage{framed}
\usepackage{placeins}

\newtheorem{theorem}{\bf Theorem}[section]

\newcommand{\etal}{{\it et~al.}\xspace}
\newcommand{\ie}{{\it i.e.}\ }
\newcommand{\eg}{{\it e.g.}\ }

\newcommand{\Ito}{It\^{o}\xspace}

\newcommand{\ave}[1]{\left\langle#1 \right\rangle}

\newcommand{\prob}{\mathbb{P}}

\newcommand{\be}{\begin{equation}}
\newcommand{\ee}{\end{equation}}
\newcommand{\bi}{\begin{itemize}}
\newcommand{\ei}{\end{itemize}}

\newcommand{\seclabel}[1]{\label{section:#1}}
\newcommand{\secref}[1]{Section~\ref{section:#1}}

\newcommand{\elabel}[1]{\label{eq:#1}}
\newcommand{\eref}[1]{(Eq.~\ref{eq:#1})}

\newcommand{\Eref}[1]{Equation~(\ref{eq:#1})}

\newcommand{\flabel}[1]{\label{fig:#1}}
\newcommand{\fref}[1]{Fig.~\ref{fig:#1}}

\newcommand{\eps}{\varepsilon}
\newcommand{\Dx}{{\Delta x}}
\newcommand{\Dt}{{\Delta t}}
\newcommand{\dt}{{\delta t}}
\newcommand{\Du}{{\Delta u}}

\newcommand{\Dv}{{\Delta v}}
\newcommand{\dv}{{\delta v}}

\newcommand{\titlehead}[1]{}
\newcommand{\address}[1]{}
\newcommand{\subject}[1]{}
\newcommand{\keywords}[1]{}
\newcommand{\corres}[1]{}
\newcommand{\email}[1]{}
\newcommand{\aucontribute}[1]{\section*{Author contributions}#1}
\newcommand{\funding}[1]{\section*{Funding}#1}

\newcommand{\conflict}[1]{\section*{Competing interests}#1}

\begin{document}

\title{The time interpretation of expected utility theory}

\author{
Ole Peters$^{1,2}$ \and Alexander Adamou$^{1}$\\
\small $^1$London Mathematical Laboratory, 8 Margravine Gardens, W6 8RH London, UK\\
\small $^2$Santa Fe Institute, 1399 Hyde Park Road, Santa Fe, NM 87501, USA}

\maketitle

\begin{abstract}
Economic models often maximise expectation values of wealth or utility. 
In non-ergodic settings, these can differ from time-averages, so that maximising expected outcomes need not maximise -- and can systematically reduce -- long-run wealth or utility.
Ergodicity economics highlights this problem and models individual agents as maximising wealth in the long run, known as growth optimality. Two instances where expected utility maximisation maps to growth optimality are known: linear utility does this for additive wealth dynamics; and logarithmic utility for multiplicative wealth dynamics. Here we show that the mapping holds more generally when the utility function coincides with the ergodicity transformation in the growth optimal model. This mapping offers a theoretical basis for choosing utility functions and suggests the testable hypothesis that wealth dynamics are predictive of risk preferences.
\end{abstract}



\section{Introduction}
\seclabel{Intro}
\subsection{Our purpose}
The present article generalises the mapping between ergodicity economics (EE) and expected utility theory (EUT). 
Ergodicity is a property of a dynamical system or stochastic process, which guarantees that expected values coincide with time averages. 
We use the term ``ergodic'' to denote only this equality, because our arguments do not require more detailed ergodic properties, see \eg the introductory discussion in \cite{UnderwoodPaillusson2025}.

EUT provides behavioural models in which agents act to maximise expected values of utility functions, $u(x)$, of their wealth, $x$, in dollars (say). EE highlights that this optimand, expected utility, $\ave{u(x)}$, is often of no relevance to a decision maker because utility may not be ergodic, and the decision maker will not actually experience its expected value. Instead, EE proposes an alternative behavioural model, in which agents maximise the growth rate of their actual wealth over long time. For succinctness, we call this behaviour ``growth optimal.''

Consider the canonical coin-toss example of EE \cite{Peters2019b}: a fair coin is tossed repeatedly, and the gambler's wealth increases by $50\%$ on heads and decreases by $40\%$ on tails. The decision problem is whether to accept or reject the gamble. A risk-neutral agent whose utility is linear in wealth, $u(x)=x$ say, will compute the expected utility arising from the gamble as $5\%$ higher than current utility: $(1.5x+0.6x)/2=1.05x$. Such an agent will always accept the gamble.

However, the stochastic process which describes utility in this instance is not ergodic: expected utility and time-average utility differ here. As the gamble is repeated, expected utility increases exponentially at $5\%$ per round, but the utility experienced by any individual gambler decreases exponentially in the long-time limit, at about $-5\%$ per round. This is a straightforward consequence of the multiplicative dynamics: multiplying the equiprobable growth factors over two consecutive rounds yields $1.5 \times 0.6=0.9$ -- a decrease of $10\%$ in two rounds, or about $5\%$ per round.

The ergodicity economist would caution against this gamble because, while luck determines utility in any finite sequence, the systematic trend over time is negative, and the time-average growth rate of utility of about $-5\%$ per round is realised by any individual gambler with probability approaching one as time passes. The positive expected value is misleading for the individual gambler, reflecting the irrelevant fact that an infinite ensemble of independent gamblers (to which the individual has no access) will see its average utility increase.

EUT optimises expected utility, a mathematical object that in non-ergodic settings is ontologically disconnected from the experience over time of any individual decision maker. Even in very simple setups like the gamble just described, nothing prevents EUT from producing behavioural protocols which opt for a guaranteed reduction of actual utility over time -- the opposite of what is desired. 

Nonetheless, a mapping between the growth optimality of EE and EUT exists. It relies on a careful treatment of growth rates and, in particular, on the identification of a function of wealth, $v(x)$, which we call the ``ergodicity transformation.'' In simple terms, $v(x)$ converts wealth, $x$, whose additive increments can (and, in general, will) be non-ergodic, to a quantity, $v$, whose additive increments have the ergodic property. This allows the time-average growth rate of wealth -- the optimand of the EE agent -- to be expressed as an expected value, namely of the rate of change of $v$. If the utility function of the EUT agent, $u(x)$, coincides (up to a linear transformation) with the ergodicity transformation, $v(x)$, for given wealth dynamics, then the EUT and EE agents will behave identically under those wealth dynamics.

Two instances of this mapping were discussed in \cite{PetersGell-Mann2016}: additive wealth dynamics have a linear ergodicity transformation, mapping the growth-optimal agent in EE to the risk-neutral linear-utility agent in EUT; while multiplicative dynamics require a logarithmic ergodicity transformation, mapping the EE agent to the canonical risk-averse EUT agent with logarithmic utility function. Our purpose there was to explain the conceptual novelty of EE, of which growth optimality is the central behavioural model. However, our restriction to two simple and familiar wealth dynamics, and two similarly simple and familiar utility functions, may have given the impression that our arguments held only for special cases. Here we formally establish a more general mapping and deliberately pick examples which are not typically studied; for instance, we work through the mapping for a risk-seeking exponential utility function in \secref{example}.

We again highlight the conceptual shift in EE -- a difference in how we model economic behaviour, both ontological and epistemological. EE highlights that growth optimality can be understood as the preference for more wealth over less in situations where the passage of time reduces uncertainty. This will be worked out formally in \secref{Technical}.

\subsection{History of decision theory}
The beginning of decision theory coincides with the beginning of formal probability theory. In 1654, Pascal and Fermat famously invented the expectation value, which led to ``expected value theory,'' where agents maximise the expectation value of their future wealth. 
The model was challenged in 1713 when Nicolas Bernoulli proposed a gamble problem, known as the St. Petersburg paradox, where future wealth is represented by a random variable with a power-law tail whose expected value does not exist. Variations of the gamble showed that real humans do not generally maximise expected future wealth, and expected value theory was considered empirically inadequate.

Modern researchers familiar with the ergodicity problem
might have questioned the validity of using an expectation value as an indication of what might happen in the future.
After all, the expectation value is not a robust statistic, and evaluating it for a stochastic process need not be representative of the process itself. But in 1713, the notion of robust statistics did not exist, and the ergodicity problem was still waiting to be discovered, which would only happen in the 1870s.

Instead of questioning the expectation operator in the computation of future expected wealth, researchers at the time focused their critique on the ability of monetary wealth, $x$, to capture what people value. In the 1720s Daniel Bernoulli and Cramer \citep{Bernoulli1738} proposed EUT.
After some clarifications, Laplace \citep{Laplace1814} described EUT in 1812 as the model where agents maximize the expectation value of an increasing non-linear transformation of wealth, the so-called ``utility function,'' $u(x)$. It implements idiosyncrasy (each agent may have a different utility function), and its non-linearity implements wealth-level dependence and risk preferences. 

Interest in a theory of choice under risk was renewed in the second half of the 20th century. 
Von Neumann and Morgenstern's work of 1944 \citep{vonNeumannMorgenstern1944}  set the tone, with an axiomatisation of Bernoulli's ideas, but without explicitly considering the ergodicity problem, which is remarkable given von Neumann's seminal contributions to ergodic theory in the early 1930s. 
Bernoulli had replaced objective monetary wealth with subjective utility, and researchers soon replaced probability and time with subjective ``felt'' or ``perceived'' probability and time. Many other developments followed, including EUT-cousins such as prospect theory and cumulative prospect theory, though apparently without much empirical success.
In 2017, following a comprehensive review of some 70 years of research, economists Friedman, Isaac, James, and Sunder arrived at the sobering assessment ``that EUT (and its cousins) fail to offer useful predictions as to what actual people end up doing.''  \cite[p.1]{FriedmanETAL2014}

From the EE perspective, this is unsurprising. The optimand at the heart of EUT and its descendants is of no \textit{a priori} relevance to the modelled individual, since it is not physically accessible to the individual. Why then would ``actual people end up doing'' what the theory predicts? 
The mistake remains ontological even if we accept the ``small-world'' assumption that all possible future events, associated probabilities, and changes in wealth are known. The expectation value of wealth or utility or any other quantity is an average across the statistical ensemble -- a collection of imagined possible worlds, only one of which will come to pass. What is missing is a means to actualise such an average, to relate it to the actual (\ie probability-one in the small-world setting) experience of the decision-maker. 

A striking way to capture the depth of the failure of EUT is the phrase: {\it expected utility maximisers don't maximise utility.} In this sense EUT fails on its own terms: an agent who maximises {\it expected} utility is guaranteed (with probability one) to lose {\it actual} utility over time. Contrast this with the EE agent, who maximizes the growth of {\it actual} wealth and therefore {\it actual} utility over time. 
The only exception is the EUT agent whose utility function is the ergodicity transformation, $u(x)=v(x)$. In this case, EUT is precisely identical to the growth optimal behaviour of EE.

Without a principled way of connecting the ensemble considered in EUT to the lived reality of decision-makers, EUT makes arbitrary predictions. The most prominent attempted remedies add extra parameters, but as Friedman \etal point out: ``Any additional parameters in a theory must pay their way in commensurate extra explanatory power [...]. EUT and its many generalisations have not yet passed this simple test in either controlled laboratory or field settings. These theories arrive endowed with a surfeit of free parameters, and sometimes provide close ex post fits to some specific sample of choice data. The problem is that the estimated parameters [...] exhibit remarkably little stability outside the context in which they are fitted. Their power to predict out-of-sample is in the poor-to-nonexistent range.''\cite[p.3]{FriedmanETAL2014}

\subsection{Ergodicity economics}
EE revisits the foundations of decision theory. Indeed, it began in earnest with a solution of the St. Petersburg paradox \citep{Peters2011b}. We ask explicitly if and under what circumstances the expectation operator of Pascal and Fermat is ontologically adequate, leaving aside the hedonic question of the pleasure derived from monetary wealth. In this way we arrive at the generalised mapping between growth optimality and EUT: agents who behave according to EUT will optimise their wealth and utility in the long run if and only if they use as the utility function what we call the ergodicity transformation.

More generally, EE questions the use of expected value and inquires about a physical rationale. Is an actual average, formed by pooling and sharing across a large and accessible sample of entities, present in our problem? Or is the observable whose expectation value is taken ergodic, with observation times long compared to mixing times, such that the expectation value has a valid temporal interpretation? By replacing unjustified expectation values with appropriate quantities, often time averages, EE resolves several puzzles in modern economic theory. For an introductory text see \citep{PetersAdamou2025}. 

Insisting on temporal optimisation for agents operating over time produces behavioural models with an evolutionary rationale: such agents will accumulate more resources than other agents in the long run, and will come to dominate their environment. This is not the case in EUT, where, as we already noted, expected utility maximisers do not maximise utility.

In EE we start by building a model of temporal wealth evolution. Such models take the form of stochastic processes, whose generating equations, known here as wealth dynamics, depend primarily on the modelled agent's economic circumstances and not their preferences. For example, an agent with labour income and no savings may be modelled as experiencing additive changes in wealth, while for an agent with significant invested capital, multiplicative dynamics may be more realistic.
As a behavioural theory, EE emphasises the power of dynamic circumstances, whereas EUT emphasises the importance of personality traits. 

It is reasonable to assume that long-term wealth maximisation was an aim in decision theory right from the beginning. After all, 17th century expected value theory optimises growth rates in an additive setting. In a multiplicative setting, the famous ``Kelly criterion'' of the 1950s \citep{Kelly1956} maximises growth, and the technique used there dates as far back as the 1870s \citep{Whitworth1870}. 

\subsection{Growth rates}
We have used the terms ``long-term wealth optimisation'' and ``growth optimisation'' synonymously. This equivalence can be seen by considering growth rates. 

What is a growth rate? In simple terms, its value is the coefficient of time in a growth function -- the $g$ in some $f(gt)$. Different dynamics correspond to different growth functions and, therefore, to different functional forms of growth rates. For example, wealth that grows linearly in time, $x=gt$, has its growth rate extracted trivially by the operation $\mathrm{d} x/\mathrm{d} t=g$; while for exponential growth, $x=\exp(gt)$, the growth rate is extracted by $\mathrm{d}\ln x/\mathrm{d} t=g$. In both cases, a constant is obtained from a time-varying function as the rate of change of a transformed wealth, linear and logarithmic in these simple examples. In stochastic growth models, growth rates fluctuate and there are no transformations of wealth whose rates of change are constant. Instead, a function, the ergodicity transformation, is chosen to extract an ergodic growth rate from the stochastic process. For a pedagogical introduction starting with deterministic dynamics, see \citep[Chapter~5.3]{PetersAdamou2025}.

As promised in \citep{PetersGell-Mann2016} and discussed briefly in \citep{Peters2019b}, we generalise growth optimality here beyond additive and multiplicative wealth dynamics. 
That is, we generalise the correspondence between EE and EUT by noting that the time average of the ergodic growth rate for a given wealth dynamic is, by virtue of its ergodicity, equal to its expectation value. This allows growth optimal decisions to be formulated as maximising expected utility, provided the utility function is the ergodicity transformation. We show that this correspondence exists for a class of wealth dynamics that admit an ergodicity transformation, and we derive the conditions for membership of this class. Going in the opposite direction, we show that any utility function with a twice-differentiable inverse has corresponding wealth dynamics for which expected utility maximisation is growth optimal.

This correspondence between two conceptually different but well established behavioural models -- growth optimality and EUT -- opens tantalising new avenues for research. Since wealth dynamics are imposed on agents, the growth-optimal utility function is known \textit{a priori} under the EE framework. In other words, EE provides a theoretical basis for the choice of utility function in EUT, absent from the latter framework. The EE model of growth optimisation is therefore falsifiable: a change in wealth dynamics should induce a change in apparent utility function. Recent experiments show robust evidence of such changes \citep{MederETAL2021}, supporting the EE prediction that dynamics are strong determinants of preferences. More broadly, EE emphasises the role of context and circumstances in decision-making, and de-emphasises the role of idiosyncratic preferences and personality. 

\section{Expected utility theory}
We provide a brief recap of Laplacian EUT. This is the theory we will use as our point of departure because it allows an exact mapping with growth optimality. We will refer to it from now on simply as EUT. It uses a cardinal utility function $u(x)$ and ignores any non-financial attendant circumstances of decisions.

For an agent facing a choice between different courses of action, the workflow of 
EUT is as follows.
\begin{enumerate}
\item Imagine everything that could happen under the different actions:\\
{\it Associate with any action $A,B,C,\dots$ a set of 
possible future events $\Omega_A,\Omega_B,\Omega_C,\dots$.}

\item Estimate how likely the consequences of each action are and how they affect your wealth:\\
{\it For set $\Omega_A$, associate a probability $p(\omega_A)$ and a change in
wealth $\Delta x_{\omega_A}$with each 
elementary event $\omega_A \in\Omega_A$, and similarly for all other sets.}

\item Specify how these outcomes affect your utility:\\
{\it Define a utility function, $u(x)$, that depends only on wealth and 
describes the agent's risk preferences.}

\item Aggregate the expected utility changes for each event:\\
{\it Compute the expected changes in utility associated with each 
available action, $\ave{\Delta u_A}=\sum_{\Omega_A} p(\omega_A) u(x+\Delta x(\omega_A)) -u(x)$, and similarly for actions $B,C,\dots$.}

\item Pick the action that increases your expected utility most:\\
{\it The action with the highest expected utility change is the agent's preferred choice.}
\end{enumerate}

We assume all steps in this workflow are possible and that all possible future events, associated probabilities, and changes in wealth are known (the small-world assumption). For simplicity, we also make the common assumption that the time between taking an action and experiencing the corresponding change in wealth is independent of the action taken. 

\section{Growth optimality as long-term wealth maximisation}
\seclabel{Technical}

We suppose that an individual's wealth evolves over time according to a stochastic process, namely a family of random numbers parametrised by time, $t$. This alone is already a departure from classical decision theory, where wealth is described by a random variable without dynamics. 
While that random variable usually models {\it future} wealth, a random variable is fundamentally an atemporal object, namely a set of pairs of numbers --- probabilities and corresponding values that the random variable may take. 
The step from random variable to stochastic process may be thought of as specifying a mode of repetition of a gamble, see \citep{PetersGell-Mann2016}.

In our setup, the individual is required to choose one from a set of alternative stochastic processes, say $x(t)$ and $x^*(t)$. We suppose here
that this is done by considering how the decision-maker would fare in the long-time limits of each process.

At each decision time, $t_0$, our individual acts to maximise 
subsequent changes in wealth by selecting $x(t)$ so that
waiting long enough under the chosen process will lead to greater wealth with certainty than under the alternative process. Mathematically speaking, there exists a sufficiently 
large $t$ such that the probability of the chosen $x(t)$ being greater
than the alternative $x^*(t)$ is arbitrarily close to one,
\begin{equation}
\forall \eps, x^*(t) \quad \exists \Dt>0 \quad \text{s.t.} \quad \prob(\Dx > \Dx^*) > 1 - \epsilon,
\elabel{max_Dx}
\end{equation}
where $0<\eps<1$ and
\begin{equation}
\Dx \equiv x(t_0 + \Dt) - x(t_0),
\elabel{Dx}
\end{equation}
with $\Dx^*$ similarly defined.

The criterion is necessarily probabilistic since the quantities $\Dx$ and 
$\Dx^*$ are random variables and it might be possible for the latter to 
exceed the former for any finite $\Dt$. Only in the limit $\Dt \to \infty$ does
the randomness vanish from the system.

Conceptually this criterion is tantamount to maximising the long-time change in wealth,
$\lim_{\Dt\to\infty}\{\Dx\}$ or, equivalently, the long-time rate of change in wealth, $\lim_{\Dt\to\infty}\{\Dx/\Dt\}$. 
However, neither of these limits is guaranteed to exist. For example, consider a 
choice between two wealth processes, $x(t)$ and $x^*(t)$, which follow geometric Brownian motion (GBM). Each process is generated by a stochastic differential equation (SDE),
\begin{eqnarray}
dx &=& x(\mu dt + \sigma dW),\\
dx^* &=& x^*(\mu^* dt + \sigma^* dW),
\end{eqnarray}
which we interpret under the \Ito convention. We will assume that $\mu > \sigma^2/2$ and $\mu^* > {\sigma^*}^2/2$ such that both wealth processes exhibit positive long-run growth (almost surely). Then, the quantities $\Dx/\Dt$ and $\Dx^*/\Dt$ diverge in the limit $\Dt\to\infty$, and a criterion requiring the larger to be selected fails to yield a decision.


To overcome this problem we introduce a monotonically 
increasing function of wealth, which we call $v(x)$. We define:
\begin{eqnarray}
\Dv &\equiv& v(x(t_0+\Dt)) - v(x(t_0));\\
\Dv^* &\equiv& v(x^*(t_0+\Dt)) - v(x^*(t_0)).
\elabel{Du}
\end{eqnarray}
The monotonicity of $v(x)$ means that the statement ``$v(x)>v(x^*)$'' is 
equivalent to the statement ``$x>x^*$.'' This implies that the events $\Dx>\Dx^*$ and 
$\Dv>\Dv^*$ are the same. 
Dividing this inequality by $\Dt$
allows us to express this event as $\Dv/\Dt>\Dv^*/\Dt$. The decision criterion in 
\eref{max_Dx} now becomes
\begin{equation}
\forall \eps, x^*(t) \quad \exists \Dt \quad \text{s.t.} \quad \prob\left(\frac{\Dv}{\Dt} > \frac{\Dv^*}{\Dt}\right) > 1 - \epsilon.
\elabel{max_Du}
\end{equation}
In words, we can say with any chosen level of certainty, $1-\eps$, 
that if we let a sufficiently long time $\Dt$ pass, the rate at which $v$ will have grown if we choose
the process $x(t)$ will exceed the rate at which it would have grown had we chosen any other available 
process.

Our decision criterion has been recast such that it focuses on the rate of change
\begin{equation}
r \equiv \frac{\Dv}{\Dt}.
\end{equation}
This motivates considering the long-time growth rate
\begin{equation}
\bar{r} \equiv \lim_{\Dt\to\infty}  \left\{ \frac{\Dv}{\Dt} \right\} =  \lim_{\Dt\to\infty} \{r\}.
\elabel{barr}
\end{equation}

However, just as $\lim_{\Dt\to\infty}\{\Dx/\Dt\}$ may be divergent, so may $\bar{r}$, and we require explicit conditions to avoid this. The long-time growth optimality criterion introduced above can be represented by a scalar objective if the following two conditions hold:
\bi
\item[(i)] Existence of a time-average growth rate.\\
There exists a random variable $\bar{r}:=\lim_{\Dt \to \infty} r$.
\item[(ii)] Ergodicity of the growth rate.\\
The observable $r$ is ergodic, $\bar{r}=\lim_{\Dt \to \infty} r=\lim_{\Dt \to \infty} \ave{r}$ (equality of averages, see \eg \cite[p.~32]{KloedenPlaten1992} or \cite{UnderwoodPaillusson2025}).
\ei

{\it Remark.}\\
Condition (i) would not be sufficient for our purposes as it allows the pathwise existence of $\bar{r}$, dependent on initial conditions, in the presence of multiple ergodic components. The limit $\Dt \to \infty$ in condition (ii) allows for possible transient behaviour of the growth rate. In the remainder of this section we restrict attention to processes with stationary growth-rate statistics, for which $\ave{r}$ is independent of $\Dt$, and the limit is redundant. We therefore simplify notation and write $\ave{r}$ without stating the limit explicitly.
Under conditions (i) and (ii), the probabilistic dominance criterion \eref{max_Du} is equivalent to choosing the process with the larger value of $\bar{r}$.

Conditions (i) and (ii) concern only the wealth process and its evolution over time. To arrive at a mapping between expected utility maximisation and growth optimisation, an additional identification is required.
\bi
\item[(iii)] Equality of ergodicity transformation and utility function.\\
The ergodicity transformation $v$ is the utility function of the decision maker, $v=u$.
\ei

This gives our central result: 
\begin{framed}
If all three conditions are satisfied, then maximising $\bar{r}$ is equivalent to maximising expected utility, and the decision theories of growth optimisation and expected utility maximisation coincide.
\end{framed}

Thus growth optimality in EE can be viewed as an axiomatisation of Laplacian EUT in dynamical terms. We offer this as an alternative to von Neumann and Morgenstern's axiomatisation of Laplacian EUT, which uses psychological terms, namely idiosyncratic preferences.

We note that $\Dv$ and hence $r$ are random variables but $\bar{r}=\ave{r}$ is not. 
Taking the long-time limit, $\lim_{\Delta t \to\infty}$, is one way of removing randomness from the problem, 
and taking the expectation value is another. The expectation value is simply
another limit: it's an average over $N$ realisations of the random number 
$r$ in the limit $\lim_{N\to\infty}$. The effect of removing randomness is that the 
process $x(t)$ is collapsed into the scalar $\ave{r}$. Complete and transitive 
preferences are the result, as is always the case when each available option 
can be associated with a number. 
Of course, the same individual will associate different numbers $\ave{\Dv}$ with the same gamble if the mode of repetition (encoded in the wealth dynamics) changes. EUT would consider such behaviour ``irrational'' or an instance of preference reversal because Laplace's setup does not take the dynamics into account. Behavioural experiments suggest that changes in dynamics are indeed accompanied by predicted changes in EUT-based valuation \cite{MederETAL2021,VanhoyweghenETAL2022}.

Our aim is to 
find pairs of wealth processes $x(t)$ and ergodicity transformations $v(x)$. 
To do so, we construct a large class of dynamics $v(t):=v(x(t))$ for which conditions (i) and (ii) hold.
To do this, we examine the long-time limit of $r$, \eref{barr}. We start by expressing the change in utility, $\Dv$, as a sum over $M$ equal time intervals,
\begin{eqnarray}
\Dv & \equiv & v(t_0+\Dt) - v(t_0) \\
& = & \sum_{m=1}^M \left[ v(t_0+m\dt) - v(t_0+(m-1)\dt) \right] \\
& = & \sum_{m=1}^M \dv_m(t),
\end{eqnarray}
where $\dt\equiv\Dt/M$ and $\dv_m(t)\equiv v(t_0+m\dt) - v(t_0+(m-1)\dt)$. From \eref{barr} we have
\begin{eqnarray}
\bar{r} & = & \lim_{\Dt\to\infty} \left\{ \frac{1}{\Dt} \sum_{m=1}^M \dv_m \right\} \elabel{barrSum}
\\
& = & \lim_{M\to\infty} \left\{ \frac{1}{M} \sum_{m=1}^M \frac{\dv_m}{\dt} \right\} \elabel{barrSum2},
\end{eqnarray}
letting $M$ diverge while keeping $\dt$ fixed.

Next, we examine the expectation value of $r$, written explicitly as an ensemble average,
\begin{equation}
\ave{r} = \lim_{N\to\infty} \left\{ \frac{1}{N} \sum_{n=1}^N \frac{\Dv_n}{\Dt} \right\},
\elabel{averSum}
\end{equation}
where each $\Dv_n$ is drawn independently from the distribution of $\Dv$, and $\Dt$ is a finite duration.

We now compare the two expressions \eref{barrSum2} and \eref{averSum}. 
Clearly the value of $\bar{r}$ in \eref{barrSum2} cannot depend on the way 
in which the diverging time period is partitioned, so the length of interval $\dt$ 
must be arbitrary and can be set to the finite duration in \eref{averSum}. 
Expressions \eref{barrSum2} and \eref{averSum} are then equivalent 
if the successive additive increments, 
$\dv_m(t)$, are distributed identically to the $\Dv_n$ in \eref{averSum}. 
This condition is satisfied if the increments are stationary and independent.

Thus we have a condition on $v(t)$ which suffices to make $\bar{r}=\ave{r}$, 
namely that it be a stochastic process whose additive increments are 
stationary and independent. This means that $v(t)$ is, in general, a L\'evy process. 
To limit the scope of the discussion to a particularly tractable class of dynamics, 
we shall restrict our attention to processes with continuous paths. According to a 
theorem stated in \cite[p.~2]{Harrison2013} and proved in \cite[Chapter~12]{Breiman1968}, 
$v(t)$ must then be a Brownian motion (BM) with drift, which obeys the \Ito SDE,
\begin{equation}
dv = \alpha dt + \beta dW.
\elabel{du}
\end{equation}
$\alpha$ and $\beta$ are constant parameters, known respectively as drift and volatility, and $dW$ is the infinitesimal increment of the Wiener process.

For BM with drift, conditions (i) and (ii) are readily verified, with $\bar{r}=\alpha$. This case will serve as our canonical example.

This is a fascinating general connection. If the physical reason 
why we observe a non-linear utility function is the non-linearity of the prevailing growth process and attendant non-linear effects of fluctuations
over time, then an observed utility function encodes a 
perceived wealth process. In other words, observed behaviour reflects the situation the decision-maker believes themself to be in. 
Provided a utility function 
$u(x)$ is invertible (\ie its inverse, $x(u)$, exists), 
a simple application of It\^o calculus to \eref{du} yields directly the 
SDE obeyed by the wealth, $x$. Thus every invertible utility 
function encodes unique wealth dynamics which arise from 
BM with drift in utility. This is explored further below.

\section{Wealth dynamics from a utility function}
We now illustrate the relationship between utility functions and wealth dynamics in the EE framework. The key idea, just discussed, is to assume that the utility function is the ergodiciy transformation, $u(x)=v(x)$ and that utility follows a BM with drift, \eref{du}. For a specified utility function, $u(x)$, we want to find the SDE followed by $x$ that is consistent with $u$ following \eref{du}. In the real world, this allows us to ask under what dynamical circumstances observed behaviour would be optimal. We may thereby discover explanatory situational constraints for a person's behaviour.

We will make liberal use of the result known as \Ito's lemma \citep{Ito1951}, which we summarise here. \Eref{du} is an \Ito SDE for $u$. \Ito's lemma says that a twice-differentiable and possibly time-dependent function of $u$, say $f(u,t)$, is also governed by an \Ito SDE, namely
\begin{equation}
df = \left(\frac{\partial f}{\partial t}+\alpha \frac{\partial f}{\partial u} + \frac{1}{2}\beta^2\frac{\partial^2 f}{\partial u^2}\right) dt + \beta \frac{\partial f}{\partial u} dW.
\elabel{df}
\end{equation}

Since we are considering EUT with fixed preferences, we confine our attention to utility functions with no explicit time dependence. If $u(x)$ can be inverted to $x(u)=u^{-1}(u)$ and if $x(u)$ is twice-differentiable, then we can substitute $f(u,t)=x(u)$ into \eref{df} to find the wealth dynamics that correspond to BM with drift in utility,
\begin{equation}
dx = \left(\alpha x'(u) + \frac{1}{2}\beta^2 x''(u)\right) dt + \beta x'(u) dW,
\elabel{dx}
\end{equation}
using Lagrange's notation for the derivatives of $x(u)$.

Note that the partial time-derivative in \eref{df} has vanished and that \eref{dx} is an autonomous SDE, \ie its coefficient functions have no explicit time-dependence. This means that the wealth dynamics do not change structurally over time. In the correspondence we are demonstrating, fixed preferences in EUT are associated with fixed dynamics in the EE framework. 

We have thus shown that 
\begin{theorem}
For any invertible utility function, $u(x)$, with twice-differentiable inverse, $x(u)$, a class of corresponding wealth dynamics, $dx$, exists such that the expected rate of change of utility is the time-average growth rate of wealth.
\end{theorem}
As a consequence, optimizing expected changes in such utility functions 
is equivalent to optimizing the time-average growth, in the sense 
of \secref{Technical}, under the
corresponding wealth process. 

\subsection{Examples}

\Eref{dx}, creates pairs of utility functions $u(x)$ and dynamics $dx$. 
These utility functions are, by construction, ergodicity transformations $u(x)=v(x)$. In discrete time, two such pairs were investigated in \citep{PetersGell-Mann2016}, repeated in \secref{lin} and \secref{log} below.

\subsubsection{Linear utility}
\seclabel{lin}
The trivial linear utility function, $u(x)=x$, corresponds to additive wealth dynamics (BM with drift),
\begin{equation}
dx=\alpha dt + \beta dW.
\elabel{dx_lin}
\end{equation}

\subsubsection{Logarithmic utility} 
\seclabel{log}
Introduced by Bernoulli in 1738 \citep{Bernoulli1738}, the logarithmic utility function, $u(x)=\ln x$, is in wide use and corresponds to multiplicative wealth dynamics, specifically GBM:
\begin{equation}
dx=x\left[\left(\alpha +\frac{1}{2} \beta^2\right) dt+\beta dW\right].
\elabel{dx_log}
\end{equation}
This case is often important in financial mathematics. 

\subsubsection{Square-root (Cramer) utility}
\seclabel{sqrt}
The first utility function ever proposed was the square-root function, $u(x)=x^{1/2}$, by Cramer in a 1728 letter to Daniel Bernoulli, partially reproduced in \citep{Bernoulli1738}. This function has a twice-differentiable inverse, $x(u)=u^2$, meaning we can apply \Ito's lemma. Substituting for $x(u)$ in \eref{dx} and carrying out the differentiations, we obtain the wealth dynamics,
\begin{align}
dx =\left(2\alpha x^{1/2} + \beta^2 \right)dt + 2 \beta x^{1/2} dW.
\elabel{dx_sqrt}
\end{align}

The drift and volatility coefficients here contain terms proportional to $x^{1/2}$, compared with constant terms in \eref{dx_lin} and terms proportional to $x$ in \eref{dx_log}. This means that the amplitude of wealth fluctuations produced by \eref{dx_sqrt} increases with wealth, unlike in additive dynamics where it is fixed, but more weakly than in multiplicative dynamics. Thus Cramer's utility function implies dynamics which lie, in some sense, between those two familiar cases. We explore this point further in \secref{iso} below.

We remark that \eref{dx_sqrt} is reminiscent of the Cox-Ingersoll-Ross model \citep{CoxIngersollRoss1985} for interest rate dynamics in financial mathematics, especially if $\alpha<0$. Similar dynamics, \ie with noise amplitude proportional to $x^{1/2}$, are also studied in absorbing-state phase transitions in statistical physics \citep{MarroDickman1999,Hinrichsen2000}.

\subsubsection{Isoelastic utility}
\seclabel{iso}

The family of \textit{isoelastic} or \textit{constant relative risk aversion} (CRRA) utility functions is a one-parameter family of functions defined in the domain $x\in (0, \infty)$ by:
\begin{equation}
u(x;\eta) \equiv
\begin{cases}
\frac{x^{1-\eta}-1}{1-\eta}, & \eta\neq1; \\
\ln x, & \eta=1.
\end{cases}
\elabel{iso}
\end{equation}
We note that the specific form $u(x;1)=\ln x$ is obtained from the general form by applying l'H\^{o}pital's rule in the $\eta\to1$ limit.

The isoelastic functions have some convenient properties for use as utility-of-wealth functions in EUT. They are monotonically increasing in $x$, consistent with the fundamental preference for more wealth to less. And they have a free parameter, $\eta$, that determines their concavity or convexity: $u(x;\eta)$ is concave in $x$ when $\eta>0$; linear when $\eta=0$; and convex when $\eta<0$. In the EUT paradigm, concavity of the utility function is interpreted psychologically as reflecting an agent's aversion to risk. Thus, by varying the $\eta$ parameter, the family of isoelastic functions can be used to model agents who are risk averse ($\eta>0$), risk neutral ($\eta=0$), or risk seeking ($\eta<0$). This flexibility has led to widespread use.

We can now interpret isoelastic utility dynamically. Since we already found the wealth dynamics corresponding to $u(x)=\ln x$ in \secref{log}, we shall focus here on the general $\eta\neq 1$ case. The function $u(x;\eta)$ is invertible,\footnote{Note that $u(x;\eta)$ is bounded from above by $1/(\eta-1)$ for $\eta>1$; and, in the domain $0\leq x<\infty$ from below by 0 for $\eta<1$. Because invertibility requires injectivity and surjectivity, the ranges have to be adjusted appropriately with $\eta$.}
\begin{equation}
x(u;\eta) = \left[(1-\eta)u+1\right]^\frac{1}{1-\eta},
\end{equation}
and twice-differentiable,
\begin{eqnarray}
&x'(u;\eta) = \left[(1-\eta)u+1\right]^\frac{\eta}{1-\eta} = x^\eta, \\
&x''(u;\eta) = \eta\left[(1-\eta)u+1\right]^\frac{2\eta-1}{1-\eta} = \eta x^{2\eta-1}.
\end{eqnarray}
Plugging these derivatives into \eref{dx} yields the wealth dynamics under which an expected isoelastic utility maximiser acts growth optimally:
\begin{equation}
dx = \left(\alpha x^\eta + \frac{1}{2}\beta^2\eta x^{2\eta-1}\right) dt + \beta x^\eta dW.
\elabel{dx_iso}
\end{equation}
 
The previous three example utility functions of $u(x)=x$, $x^{1/2}$, and $\ln x$ are, up to a linear transformation, special cases of the isoelastic function with $\eta=0$, $1/2$, and $1$ respectively. Thus we see that Cramer's square-root function sits between the linear and logarithmic functions in concavity (or implied risk aversion). Inserting these $\eta$ values into \eref{dx_iso} yields the same wealth dynamics we found previously (up to a constant multiplier). In effect, \eref{dx_iso} defines a family of wealth dynamics which, in the EE framework, might be used to model more general forms of real wealth evolution than linear and exponential growth, or to control wealth evolution in laboratory tests of human decisions, \eg generalising \citep{MederETAL2021}.

\subsection{Graphical interpretation}
We end this section with a graphical interpretation of the mapping between utility functions and wealth dynamics. Consider a plot of a general utility function, $u(x)$, with $x$ on the horizontal axis and $u$ on the vertical axis. We assume only that $u(x)$ increases monotonically and somewhat smoothly (specifically, its inverse must be twice-differentiable). The ergodicity constraint, $u(x)=v(x)$, from growth optimality is satisfied if $u$ executes a BM with drift, given by \eref{du}. This is a continuous-time random walk in which, for a fixed time step $\Dt$, the random utility step is drawn from a fixed distribution, $\Du\sim\mathcal{N}(\alpha\Dt, \beta\Dt)$. Graphically, we imagine utility executing a random walk on the vertical axis, where the random steps have fixed amplitude. Any bounds on $u(x)$ can be implemented as appropriate boundary conditions which leave the trajectories as genuine BM realisations, see \secref{wealth_distribution} for an example.

How does this translate into wealth steps on the horizontal axis? Trivially, if $u(x)$ is linear, then wealth steps are equal to utility steps up to a constant multiplier. This is consistent with our knowledge that wealth must itself follow BM with drift in this case. Things get more interesting when $u(x)$ is concave or convex. If concave, then the amplitude of wealth fluctuations increases as wealth increases because the slope of $u(x)$ decreases with $x$. We show this graphically in \fref{concave_convex}, left panel. This implies dynamics under which wealth tends to create (or destroy) more of itself, which we may find economically realistic.
Low wealth, under these conditions, is `sticky': more steps are required to get away from it. 

\begin{figure}[h]
  \begin{minipage}{0.48\textwidth}
    \centering
    \includegraphics[width=\textwidth]{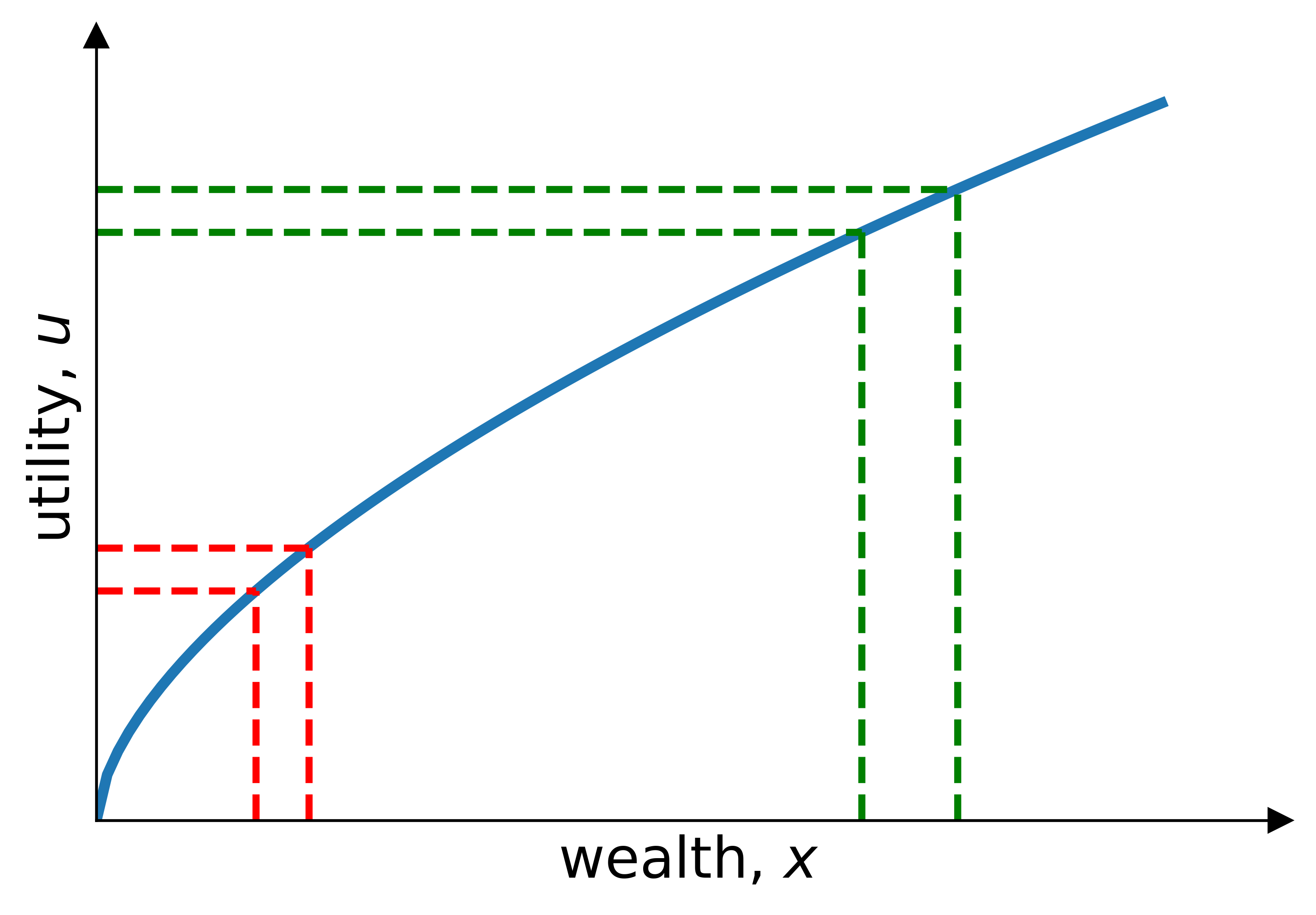}
  \end{minipage}\hfill
  \begin{minipage}{0.48\textwidth}
    \centering
    \includegraphics[width=\textwidth]{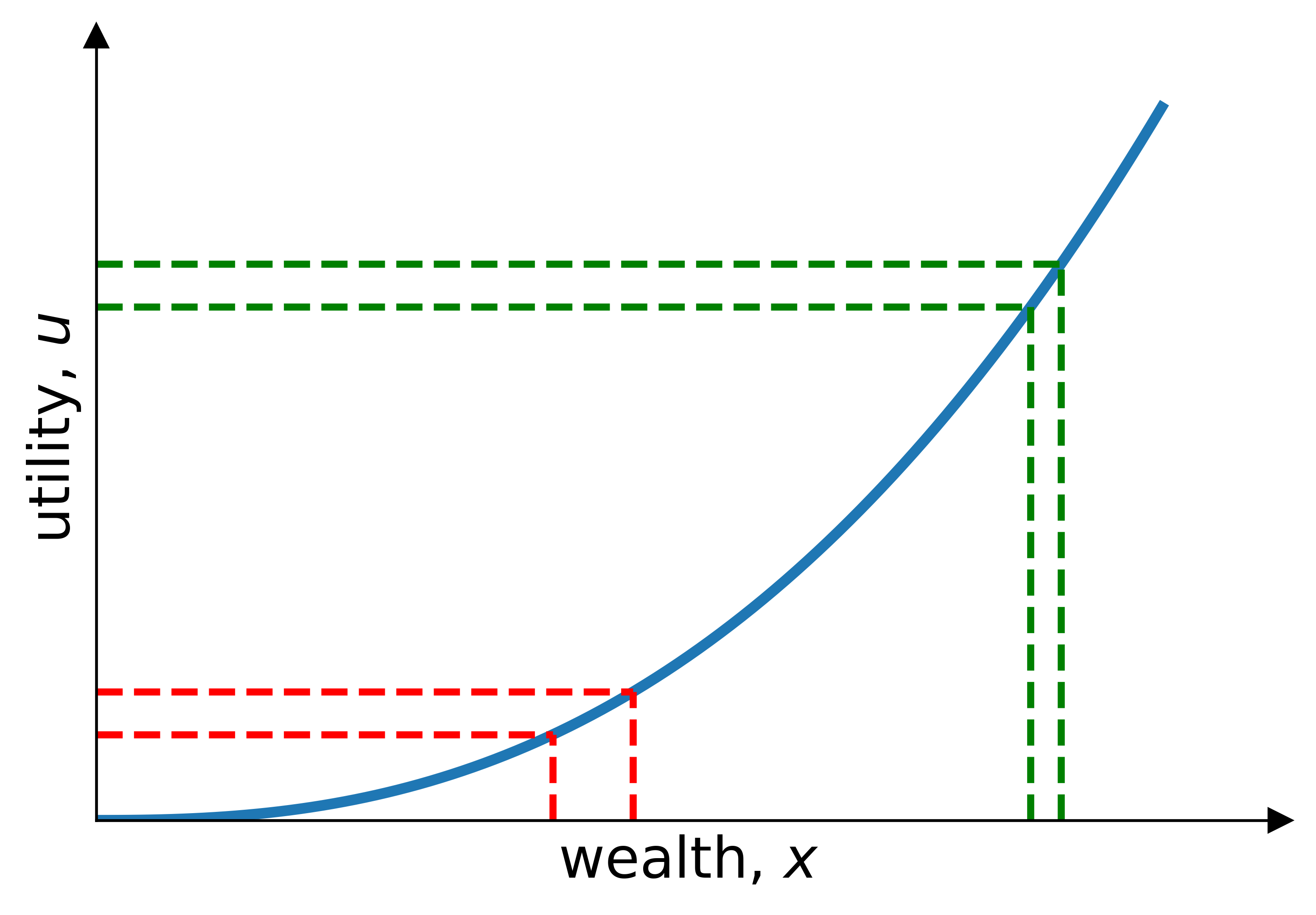}
  \end{minipage}
  \caption{{\bf Left}: With a concave utility function (blue), a fixed-sized step in utility produces a smaller step in wealth when wealth is small (red dashed) than when wealth is large (green dashed). This reflects common economic situations in which larger wealths are exposed to larger-amplitude fluctuations.
{\bf Right}: With a convex utility function (blue), a fixed-sized step in utility produces a larger step in wealth when wealth is small (red dashed) than when wealth is large (green dashed). Under the unusual dynamics implied by convex utility, larger wealths tend to be stable, since they are subject to smaller fluctuations.}
  \label{fig:concave_convex}
\end{figure}

If, on the other hand, $u(x)$ is convex, then the converse is true: the amplitude of wealth fluctuations will decrease as wealth increases, see \fref{concave_convex}, right panel. Under such dynamics, large wealth would be `sticky' and poor agents would have a strong incentive to take risks to obtain wealth, safe in the knowledge that, once they do, it will be difficult to lose. In the EUT paradigm, a convex utility function implies precisely such risk-seeking behaviour, of which the right panel of \fref{concave_convex} provides a dynamical explanation.

\section{Utility function from wealth dynamics}
\seclabel{Utility_function}
We now ask under what circumstances the procedure in
\eref{dx} can be inverted, \ie when can a utility function be found for
given wealth  dynamics? In other words, what conditions must the wealth dynamics, $dx$, satisfy so that growth optimality is achieved by maximising expected changes in some utility, $u(x)$? Within our \Ito framework, this boils down to whether a $u(x)$ exists such that the corresponding utility dynamics, $du$, give BM with drift, \eref{du}.

Suppose the wealth dynamics are described by a general autonomous \Ito SDE,
\begin{equation}
dx=a(x) dt +b(x) dW,
\elabel{dx_1}
\end{equation}
in which the coefficients $a(x)$ and $b(x)$ are explicitly functions of wealth only, which we assume to be known. We apply \Ito's lemma to \eref{dx_1} to find the equivalent SDE for $u(x)$, which we insist be the SDE for BM with drift, \eref{du}:
\begin{equation}
du = \underbrace{\left(a(x) u'(x) + \frac{1}{2}b(x)^2 u''(x)\right)}_{\alpha} dt + \underbrace{\vphantom{\left(\frac{1}{2}\right)} b(x) u'(x)}_{\beta} dW.
\elabel{du_2}
\end{equation}
Requiring this SDE to have constant coefficients yields two equations,
\begin{eqnarray}
a(x) u'(x) + \frac{1}{2} b(x)^2 u''(x) &=& \alpha, \elabel{A} \\
b(x) u'(x) &=& \beta, \elabel{beta}
\end{eqnarray}
which must be satisfied by the known functions $a(x)$ and $b(x)$ for the unknown utility function $u(x)$ and constants $\alpha$ and $\beta$.

It follows from \eref{beta} that
\begin{equation}
u'(x) = \frac{\beta}{b(x)}
\elabel{v_dash}
\end{equation}
and, therefore, that
\begin{equation}
u''(x) = -\frac{\beta b'(x)}{b(x)^2}.
\end{equation}
Substituting these expressions into \eref{A} yields an equation connecting $a(x)$ and $b(x)$,
\begin{equation}
a(x) =b(x)\left(\frac{\alpha}{\beta}+ \frac{b'(x)}{2}\right).
\elabel{consistency}
\end{equation}

\Eref{consistency} is a consistency condition on the wealth dynamics. A utility function $u(x)$ exists which maps the \Ito process for $x$ to a BM with drift if and only if the coefficient functions $a(x)$ and $b(x)$ satisfy \eref{consistency}. We do not need to find this $u(x)$ explicitly to know whether or not it exists. We need only to plug $a(x)$ and $b(x)$ into \eref{consistency}.

If the utility function does exist, the simplest way to find it is to integrate \eref{v_dash} directly,
\begin{equation}
u(x) = \int^x \frac{\beta}{b(\xi)} d\xi + C,
\elabel{v_int}
\end{equation}
where $\xi$ is a dummy variable and $C$ is a constant of integration. This constant is arbitrary and reflects the fact that only changes in utility are meaningful, as pointed out in \citep{vonNeumannMorgenstern1944}.

In the preceding calculations we assumed that the wealth dynamics are known, \ie that we know the drift and noise coefficients in \eref{dx_1}. With this knowledge, we can say whether a utility function $u(x)$ exists that is consistent with these coefficients (and, if so, what it is). More generally, the drift coefficient, noise coefficient, and utility function form an interdependent triplet: if we know two of them, we can find a consistency condition for the third.

\subsection{Example: exponential utility}
\seclabel{example}
Consider the following wealth dynamics,
\begin{equation}
dx=\left(\frac{\alpha}{\beta}e^{-x}-\frac{1}{2}e^{-2x}\right)dt+e^{-x}dW.
\elabel{test_dyn}
\end{equation}
We note that $a(x)=\frac{\alpha}{\beta}e^{-x}-\frac{1}{2}e^{-2x}$ and $b(x)=e^{-x}$.
\Eref{consistency} imposes conditions on the drift term $a(x)$ in terms of the 
noise term $b(x)$. Substituting in \eref{consistency} reveals that the consistency 
condition is satisfied by the dynamics described by \eref{test_dyn}.

A typical trajectory of \eref{test_dyn} is shown in \fref{test_dyn}, right panel (green line).

\begin{figure}[h]
  \begin{minipage}{0.48\textwidth}
    \centering
    \includegraphics[width=\textwidth]{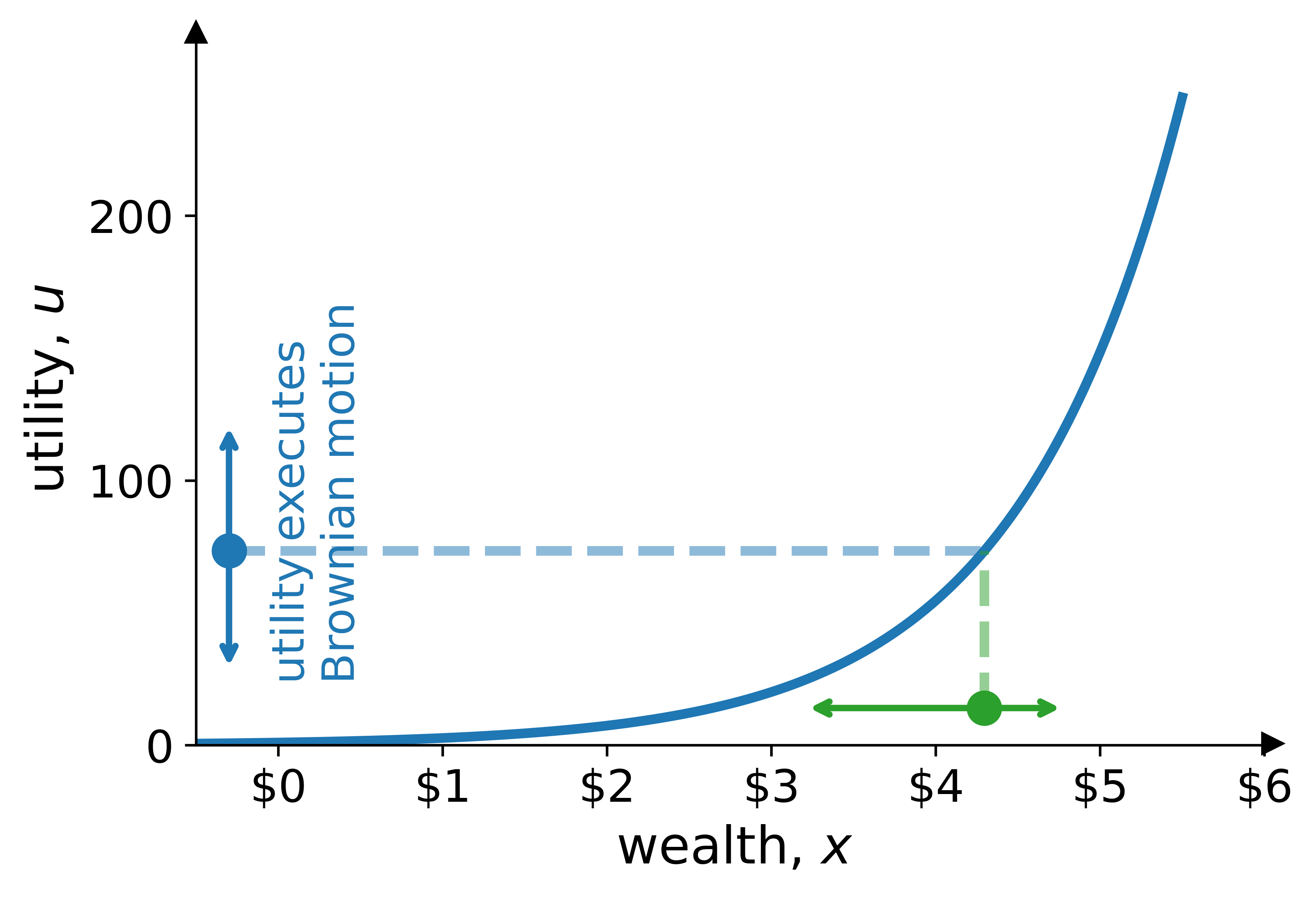}
  \end{minipage}\hfill
  \begin{minipage}{0.52\textwidth}
    \centering
    \includegraphics[width=\textwidth]{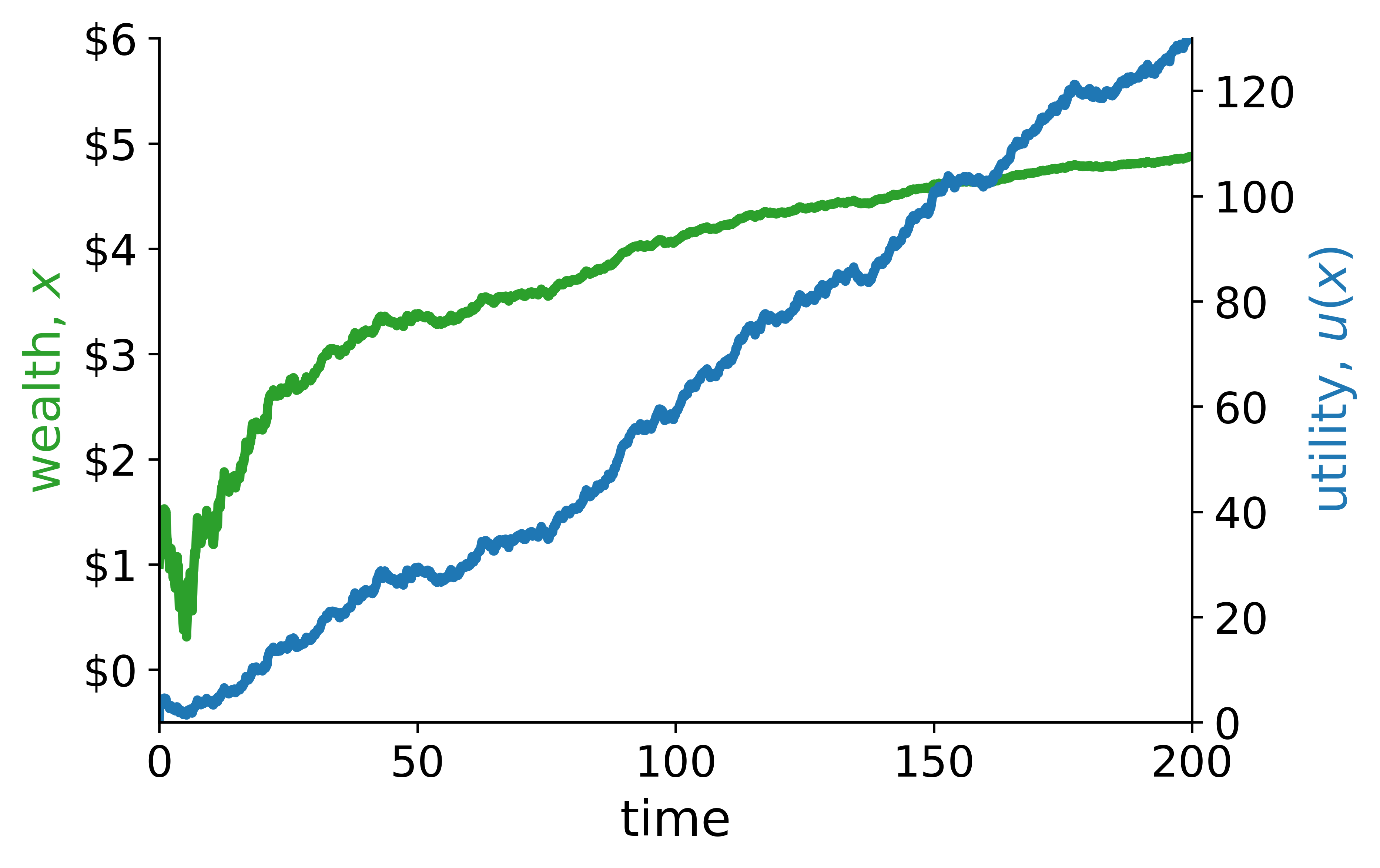}
  \end{minipage}
  \caption{{\bf Left}: Utility function \eref{test_dyn_u} corresponding to the wealth dynamic \eref{test_dyn}, using $\beta=1$ and $C=0$. If utility performs Brownian motion with drift (imagine moving the blue dot up and down), wealth will perform the corresponding dynamic (imagine moving the green dot left to right).
  {\bf Right}: Unfolding the image on the left in time, we show a wealth trajectory, $x(t)$, described by \eref{test_dyn} (green) alongside the corresponding utility trajectory (blue), which executes Brownian motion with drift, using $\alpha=0.5$, $\beta=1$, and $x_0=\$1$. 
  Because the utility function is convex, wealth growth slows down over time, and the fluctuations in wealth diminish with increasing wealth. Classical decision theory would label individuals with convex utility ``risk seeking.'' Given the dynamic in \eref{test_dyn}, the corresponding behaviour is growth optimal.}
  \label{fig:test_dyn}
\end{figure}

Because \eref{test_dyn} is internally consistent, it is possible to derive the corresponding utility function by integration as indicated in \eref{v_int}.

Substituting $b(x)=e^{-x}$ from \eref{test_dyn} in \eref{v_int} shows this
integral to be
\begin{equation}
u(x)=\int^{x}d\xi\beta e^\xi = \beta e^x +C,
\elabel{test_dyn_u}
\end{equation}
plotted in \fref{test_dyn}, left panel. This exponential utility function is monotonic and invertible, which is 
reflected in the fact that the consistency condition is satisfied. 
The utility function is convex. From the 
perspective of EUT an individual behaving optimally according to 
this function would be labelled ``risk-seeking.'' 
The dynamical perspective corresponds to a qualitatively different interpretation: 
under the dynamics in \eref{test_dyn}, appropriately ``risk-seeking'' individuals behave optimally, in the sense that their wealth will grow faster than that of risk-averse individuals. The dynamics in \eref{test_dyn} have the feature that fluctuations in wealth shrink as wealth grows. High wealth is therefore sticky -- if an individual fluctuates out of low wealth and into higher wealth, they will tend to stay there. 

\section{Wealth distribution from dynamics}
\seclabel{wealth_distribution}
The dynamical interpretation of EUT makes it possible to compute 
wealth distributions. 
We know that the dynamic of $u(t)$ is Brownian motion with drift. To determine its distribution, we need to impose appropriate initial and boundary conditions, in effect complementing local information about how individual utility paths evolve over short time steps with global information about which utility paths are admissible in our model. Having derived the distribution of $u$, we find the distribution of $x$ via the relevant change of variables.

We illustrate this using our example of exponential utility, \secref{example}. Here wealth follows \eref{test_dyn} without restriction, $x\in \mathbb{R}$. 
Considering the limits $x\to\pm\infty$ in $u(x)=\beta e^x+C$, we deduce that $u$ is restricted to $(C,\infty)$. 
Thus only the utility paths which remain above $C$ correspond to wealth paths that are admissible in our model.
We implement this restriction by conditioning: we select from the ensemble of BM trajectories only those for which $u>C$ for all $t$. Finding the probability density function (PDF) of $u$ is then 
a well-known problem with a well-known solution.

First, one finds the density of surviving paths in the full ensemble, the so-called absorption kernel, which can be derived using the method of images as
\be
P_u^{\text{abs}}(u, t;u_0)=\phi(u; u_0+\alpha t,\beta^2 t)
-e^{\frac{2\alpha(C-u_0)}{\beta^2}} \phi(u; 2C-u_0+\alpha t,\beta^2 t).
\ee
Here $\phi(u;\mu,\sigma^2)$ is the normal PDF with mean $\mu$ and variance $\sigma^2$, and $u_0>C$ is the initial utility at $t=0$.
Integrating over $u$ yields the survival probability, \ie the probability of $u$ not hitting $C$ before time $t$, as
\be
S(t;u_0)=\int_C^{\infty} du P_u^{\text{abs}}(u, t;u_0)=
\Phi\left(\frac{u_0-C+\alpha t}{\beta\sqrt{t}}\right)
-e^{\frac{2\alpha(C-u_0)}{\beta^2}}\Phi\left(\frac{C-u_0+\alpha t}{\beta\sqrt{t}}\right),
\ee
where $\Phi(u)$ is the standard normal cumulative density function (CDF).
Combining the two yields the conditioned PDF for $u$, properly normalized on $(C, \infty)$, as
\be
P_u(u,t; u_0)=\frac{P_u^{\text{abs}}(u, t;u_0)}{S(t;u_0)}.
\ee
Using $u=\beta e^x+C$ and $du/dx=\beta e^x$, this yields the PDF for $x$, which is the wealth distribution we were seeking,
\be
P_x(x,t;x_0)=\beta e^x P_u(\beta e^x+C,t;\beta e^{x_0}+C).
\elabel{wealth_distribution}
\ee

This is shown in \fref{x_dist}, right panel. The distribution is sensible given what we know about the dynamics -- since fluctuations diminish with increasing wealth many individuals will be found at high wealth (all those that have fluctuated away from low wealth), with a heavy tail towards lower wealth.

\begin{figure}[h]
  \begin{minipage}{0.48\textwidth}
    \centering
    \includegraphics[width=\textwidth]{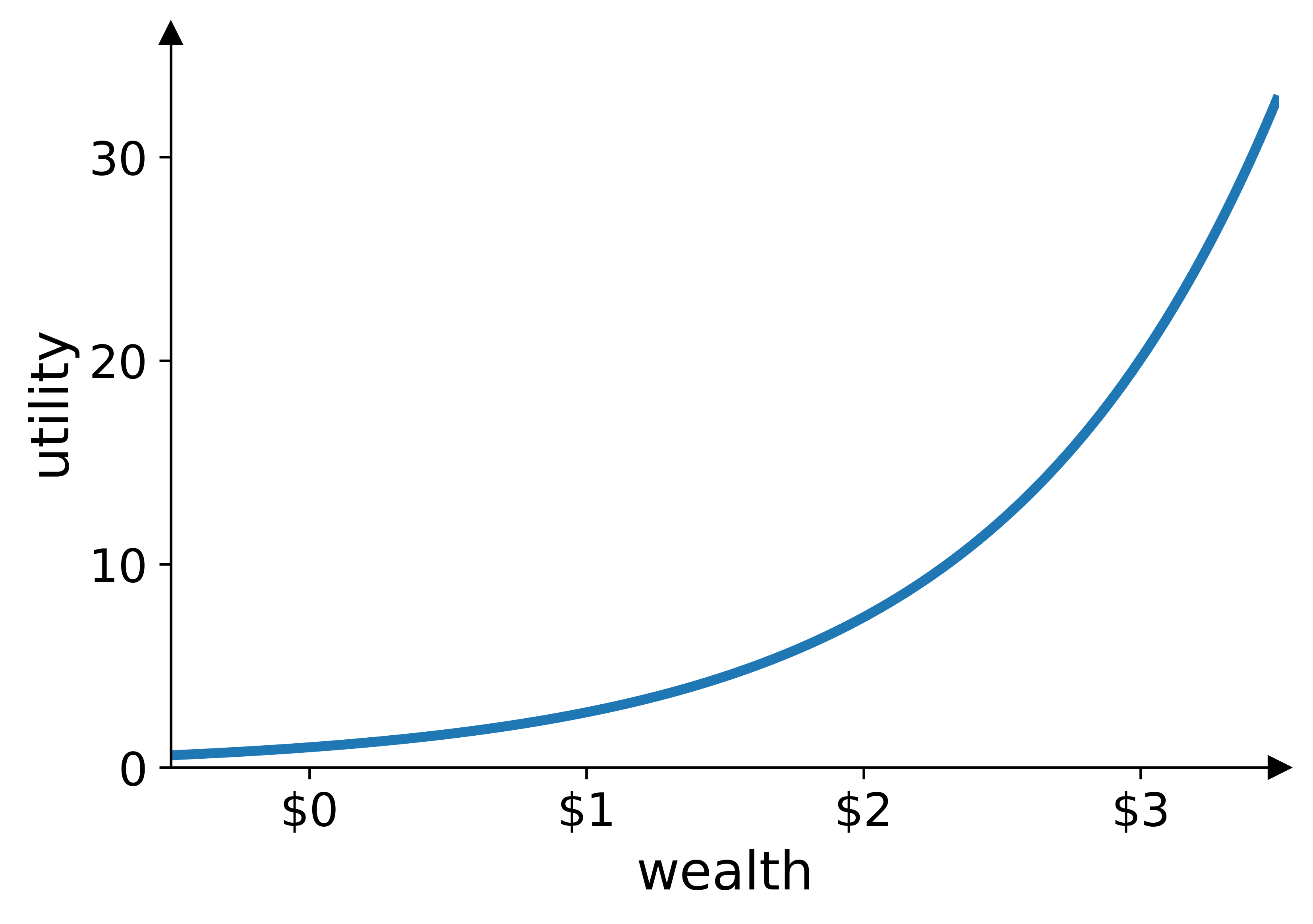}
  \end{minipage}\hfill
  \begin{minipage}{0.52\textwidth}
    \centering
    \includegraphics[width=\textwidth]{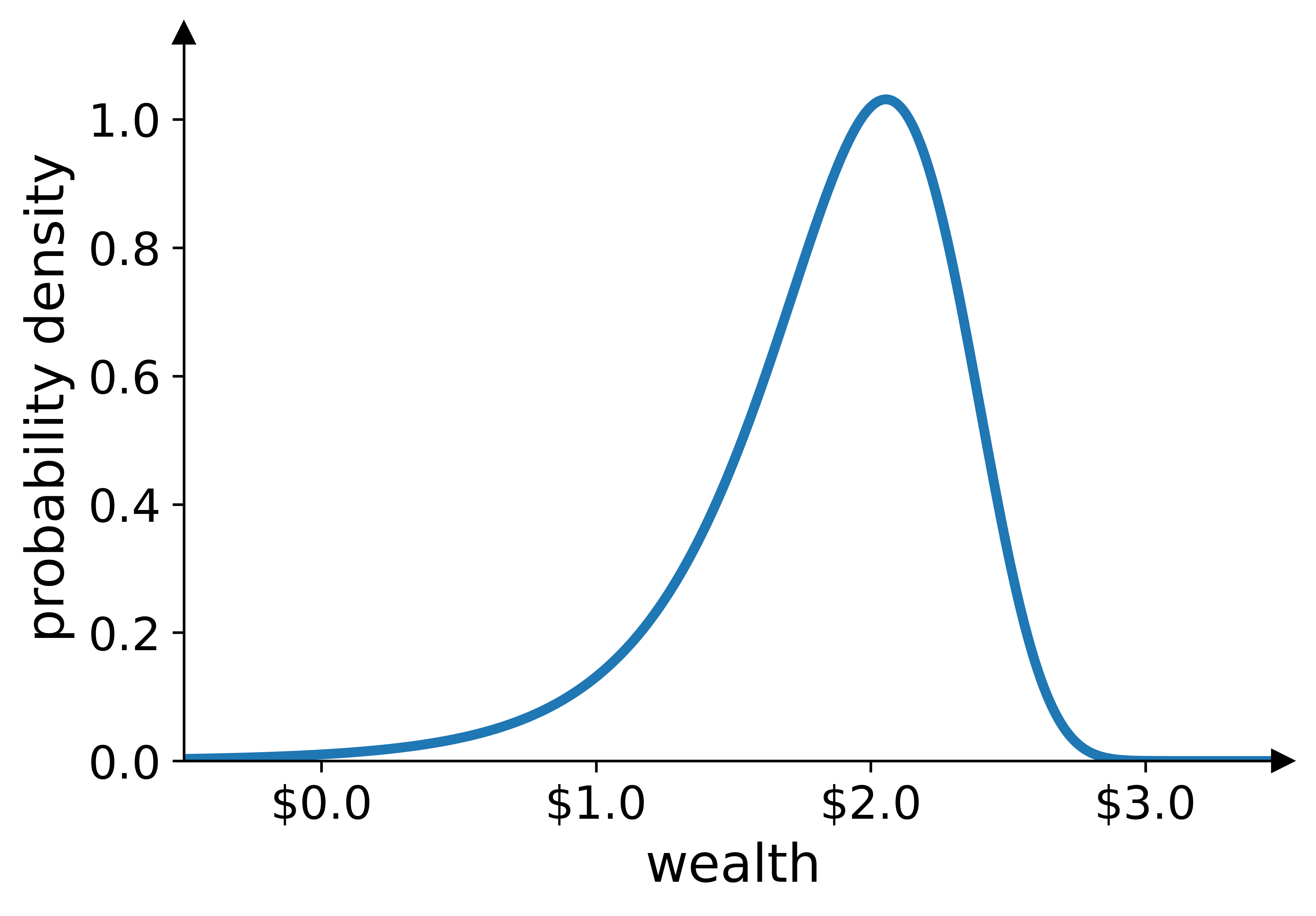}
  \end{minipage}
  \caption{{\bf Left}: Utility function of \eref{test_dyn_u}, with $\beta=1$ and $C=0$. Optimizing the expected change in this utility function also optimizes time-average growth under the corresponding dynamics, \eref{test_dyn}. An unusual utility function -- like the convex function shown here -- reflects unusual dynamics, see text.
  {\bf Right}: Probability density function of wealth, also known as the wealth distribution, \eref{wealth_distribution}. This distribution is generated by the wealth dynamics in \eref{test_dyn}. 
The time is fixed to $t=10$, and we use $\alpha=1/2$, $\beta=1$, $C=0$, and $x_0=\$1$, consistent with \fref{test_dyn}.}
  \flabel{x_dist}
\end{figure}

\section{Discussion}

EE provides a common framework for decision-theoretic problems that are traditionally treated separately, addressing: classical paradoxes such as the St Petersburg problem \cite{Peters2011b}; modern issues in asset pricing, including the equity premium puzzle \cite[Chap.10]{PetersAdamou2025}; and core questions that arise even in the absence of risk, like intertemporal choice and discounting \cite{AdamouETAL2021}. It yields insights into macroeconomics, for instance by providing formal models for the formation of wealth distributions and inequality \cite{BermanPetersAdamou2021}. It shows formally how insurance \cite{Peters2023} and cooperative social structures \cite{PetersAdamou2022} emerge as growth-enhancing strategies, resembling adaptive coin-tossing schemes put forward in evolutionary biology \cite{CooperKaplan1982}. In each case, the unifying element is broken ergodicity: a dynamic for which expected values fail to describe the long-run experience of individual agents.

Conceptually, this perspective has a long history, going back to the early days of probability theory, with  Cardano cautioning in the 16th century ``when the die once falls unlucky for him, his loss may be greater than the net gain of many wins'' \cite[p.8]{Cardano1564}. Whitworth \cite{Whitworth1870} and Kelly \cite{Kelly1956} both pointed out the inadequacy of expected-value maximisation in multiplicatively repeated gambles. Lopes \cite{Lopes1981} simulated trajectories of wealth in repeated exposures to the St Petersburg lottery, showing numerically that performance decouples systematically from expected value as the game continues. EE traces the power of these contributions to the formal ergodicity problem first formulated by Boltzmann \cite{Cohen1997}, and it helps clarify the relationship to a mainstream decision theory which relies heavily on expected value.

As a perspective on behaviour, EE emphasises effects of context and environment over effects of personal preferences. 
Using Simon's scissor metaphor: behaviour can only be understood through the joint action of two blades -- the cognitive characteristics of the agent and the structure of the environment \cite[p.7]{Simon1990}. Orthodox economics, with its roots in 17th and 18th century mathematics, tends to ascribe behavioural patterns to personality types. EE elucidates the role of the dynamic environment: agents who don't align their utility functions with the dynamics will obtain less wealth and utility than agents who do.
In this sense, EE aligns naturally with Simon's notion of ecological rationality: rationality is not an intrinsic property of the agent alone, but of the agent embedded in a particular environment.

Formally, EUT is an 18th century patch to a 17th century model of expected wealth maximisation, 
applied because the latter made blatantly wrong predictions of human behaviour. 
Since the mathematics of randomness was in its infancy in the 18th 
century, the conceptual problem of ergodicity was overlooked. Maximising expected utility does not generally maximize wealth or utility -- it does so only when the utility function coincides with the ergodicity transformation. Thus, utility theory set
economics off in a problematic direction. 
EE, on the other hand, gives a physical meaning to the non-linear
mappings people seem to apply to monetary amounts. These apparent mappings 
can be interpreted as encoding the non-linearity of the dynamics to which their wealth is subjected.

\aucontribute{O.P. and A.A. conceived the study, developed the theoretical
framework, and contributed to the analysis and writing. Both authors approved
the final manuscript.}

\funding{This work was carried out at the London Mathematical Laboratory and
was supported by philanthropic donations.}


\conflict{The authors declare that they have no competing interests.}

\bibliographystyle{abbrv}
\bibliography{bibliography}
\end{document}